\pdfoutput=1
\documentclass[superscriptaddress,reprint,floatfix]{revtex4-2}

\usepackage{amsmath}
\usepackage{amstext}
\usepackage{amssymb}
\usepackage{booktabs}
\usepackage{array}
\usepackage{graphicx}
\usepackage{epstopdf}
\usepackage{float}
\usepackage{color}
\usepackage{xcolor}
\usepackage{verbatim}
\newif\ifshowcitationbackrefs
\showcitationbackrefsfalse
\ifshowcitationbackrefs
  \usepackage[pagebackref=true]{hyperref}
\else
  \usepackage{hyperref}
\fi
\hypersetup{
  colorlinks=true,
  linkcolor=blue,
  citecolor=blue,
  urlcolor=blue,
}
\ifshowcitationbackrefs
  \renewcommand*{\backrefalt}[4]{%
\ifcase #1\relax
    \or [cited on p.~#2]%
    \else [cited on pp.~#2]%
\fi
}
\fi
\providecommand{\doi}[1]{\href{https://doi.org/#1}{doi:#1}}
\definecolor{orcidgreen}{HTML}{A6CE39}
\newcommand{\orcidlink}[1]{\href{https://orcid.org/#1}{\textsuperscript{\textcolor{orcidgreen}{iD}}}}

\newcommand{\KFE}{Korea Institute of Fusion Energy, Daejeon 34133, Republic of Korea}
\newcommand{\UST}{University of Science and Technology, Daejeon 34129, Republic of Korea}
\newcommand{\IPP}{Max-Planck-Institut f\"ur Plasmaphysik, Greifswald 17491, Germany}

\begin{document}

\title{Automated Outlier-Robust Bayesian Profile Fitting for Magnetically Confined Plasmas with Modified Tanh Profiles and Good-and-Bad Gaussian Mixture Likelihoods}

\author{Jaewook~Kim\,\orcidlink{0000-0003-0047-5498}}
\email[]{ijwkim@kfe.re.kr}
\affiliation{\KFE}
\affiliation{\UST}
\author{Jekil~Lee\,\orcidlink{0000-0002-1252-7075}}
\affiliation{\KFE}
\author{Laurent~Jung}
\affiliation{\KFE}
\author{Sang-hee~Hahn\,\orcidlink{0000-0001-8115-9248}}
\affiliation{\KFE}
\author{Sehyun~Kwak\,\orcidlink{0000-0001-7874-7575}}
\affiliation{\IPP}

\date{\today}
\begin{abstract}
We present an outlier-robust Bayesian approach for automated kinetic profile fitting in magnetically confined plasmas with the modified tanh (mtanh) parametrisation and demonstrate its implementation on KSTAR. The method addresses two systematic obstacles: anomalous diagnostic channels can bias least-squares fits, and multimodality of the mtanh cost surface can trap deterministic optimisers in secondary minima. The deployed workflow uses a good-and-bad Gaussian mixture likelihood based on the Box--Tiao formulation as the default outlier-robust likelihood for fitted diagnostic channels, with posterior outlier probabilities retained as channel-level quality indicators. The posterior is sampled with an affine-invariant ensemble MCMC sampler initialised near the result of deterministic maximum a posteriori (MAP)-seeking optimisation, reducing sensitivity to secondary minima on the multimodal mtanh surface. A batch automation layer retrieves diagnostic data from MDSplus and fits arbitrary time slices in parallel for the quantities \(n_e\), \(T_e\), \(T_i\), and \(v_T\) for which the relevant diagnostics are available. Results are written in formats suitable for MDSplus upload and downstream analysis. Representative KSTAR H-mode cases show that the mixture likelihood downweights contaminated measurements while preserving plausible pedestal profiles. The workflow provides a practical basis for future large-scale kinetic profile production for kinetic-EFIT, TRANSP, FASTRAN, and data-driven analysis workflows.
\end{abstract}

\maketitle

\vspace{2pc}

\section{Introduction}\label{sec:introduction}

Kinetic profiles of electron density \(n_e\), electron temperature \(T_e\), ion temperature \(T_i\), and toroidal rotation velocity \(v_T\) underpin almost every quantitative analysis carried out on a tokamak: kinetic-EFIT reconstructions based on EFIT equilibrium reconstruction \cite{lao2005efit}, power-balance and transport analyses with integrated modelling codes such as TRANSP \cite{pankin2025transp} and FASTRAN \cite{park2018fastran}, and magnetohydrodynamic (MHD) stability calculations including pedestal-stability evaluations against the edge pedestal (EPED) model \cite{snyder2011eped}. Producing these profiles requires combining measurements from several diagnostics with different spatial sampling geometries, resolutions, systematic errors, and, most importantly, failure modes.

For KSTAR \cite{lee2001kstar}, the relevant diagnostics include the Thomson Scattering (TS) system \cite{lee2010ts} for \(n_e\) and \(T_e\), the Two-Color Interferometer (TCI) \cite{lee2016tci} for the line integrated electron density along five tangential chords, and Charge Exchange Spectroscopy (CES) \cite{ko2010ces} for \(T_i\) and \(v_T\). Each of these systems produces, at every laser firing or beam blip, tens of channel-resolved measurements with associated error bars. Even after standard quality flags are applied, individual channels can fail in ways that are not captured by the quoted statistical uncertainty. Representative examples include stray laser light contaminating a TS chord, weak charge-exchange signals in CES channels due to beam attenuation or low impurity content, fringe jumps in a TCI line, beam blip timing problems, or hardware glitches that escape the per-shot calibration.

In a least-squares fit, each quoted error bar is treated as exact. As Sivia emphasised \cite{sivia1996duff}, one or two such ``duff'' data points (anomalous measurements whose true uncertainty is much larger than the quoted error bar) can severely skew the inferred profile and the derived pedestal parameters. This can occur even when visual inspection can often identify them as inconsistent with the bulk of the measurements. Manual masking is the customary remedy, but it does not scale to the thousands of time slices contained in a modern campaign database. Its subjective nature also harms reproducibility.

A second, less frequently discussed, obstacle to automation is the local minimum problem. The mtanh profile model is a seven-parameter nonlinear function of the flux coordinate, and its cost surface is generally non-convex in the neighbourhood of the pedestal width \(\Delta_{\rm ped}\), pedestal height \(f_{\rm ped}\), and radial shift \(\Delta x\), giving rise to a multimodal landscape with multiple local minima separated by narrow ridges.

When the edge-channel spacing is not sufficient to resolve the pedestal gradient scale, several combinations of pedestal position, width, and core shape can explain the measured points with comparable cost. A gradient-based or simplex optimiser can then converge to one such local basin, producing a fit that satisfies the optimiser stopping criteria but yields pedestal gradients that differ substantially from other plausible solutions. This failure mode can be difficult to catch in unattended operation and can propagate into downstream analysis that consumes the fitted profile, including kinetic-EFIT reconstructions and transport analyses whose inferred quantities are sensitive to local profile gradients.

A Bayesian approach via Markov-chain Monte Carlo (MCMC) sampling mitigates both obstacles simultaneously. Given sufficient run length and adequate inter-mode mixing, the chain can explore multiple posterior basins. The posterior mean and credible intervals can then reflect a broader view of the sampled landscape rather than the single nearest optimum. At the same time, the likelihood can be generalised from the least-squares Gaussian to a form tolerant to outliers. Contaminated data points are then downweighted automatically without manual masking.

This requirement is important because future KSTAR analysis workflows increasingly require kinetic profiles to be produced automatically over many shots and many time slices. Kinetic-EFIT reconstructions, TRANSP integrated-modelling runs, FASTRAN scenario modelling runs, and systematic transport or pedestal studies all depend on reproducible profile inputs that can be generated without manual tuning of every time slice. The same requirement also arises in data-driven applications: constructing data sets ready for machine learning requires consistent kinetic profile values, uncertainty estimates, and data quality indicators over large collections of discharges. Over such data sets, even infrequent failed fits can accumulate and bias aggregate conclusions. Automating the profile fitting step with built-in robustness against both outliers and local minima is therefore an enabling requirement for large-scale, reproducible analyses.

In this paper we describe an automated Bayesian profile fitting approach that mitigates both the outlier problem and the local minimum problem within a single probabilistic formulation. The deployed probabilistic interface uses a Gaussian ``good-and-bad'' mixture in the Box--Tiao formulation \cite{box1968outlier} as the default likelihood for fitted diagnostic channels. The mixture form uses configurable outlier fraction and width parameters and yields posterior outlier probabilities for each channel as a by-product. The standard diagonal Gaussian likelihood and the Sivia conservative likelihood \cite{sivia1996duff} are retained as options selectable by users for sensitivity studies.

This treatment differs from recent KSTAR work that used support vector machine regression (SVMR) to identify outliers before applying Gaussian process (GP) regression to the cleaned data set \cite{kim2024svmrgpr}. Here, enabled measurements remain in a single probabilistic fit, and discordant channels are downweighted by the mixture likelihood while posterior outlier probabilities are reported together with the fitted mtanh profile.

Mixture likelihoods have also been used in fusion plasma diagnostic inference in a different context. Krychowiak et al. used a Bayesian mixture treatment for effective charge inference from spectroscopic bremsstrahlung measurements \cite{krychowiak2004zeff}, and Kwak et al. applied the Box--Tiao good-and-bad form to handle line emission contamination as outliers in Bayesian \(Z_{\rm eff}\) inference \cite{kwak2026zeff}. To our knowledge, however, the use of a good-and-bad Gaussian mixture as the default likelihood for automated kinetic profile fitting, producing per-channel posterior outlier probabilities as a routine output, has not previously been reported.

The sampling workflow reduces sensitivity to secondary minima by exploring the posterior with an affine-invariant ensemble sampler initialised from a deterministic warm start obtained by minimising the negative log posterior. This point is the result of an optimisation attempt to locate a MAP point, not a guarantee that the global posterior maximum has been found. The inferred profile can then reflect a broader region of the sampled posterior landscape rather than committing to the single nearest optimum of a deterministic optimiser. This distinction is critical for pedestal parameters on the multimodal mtanh cost surface, where uncertainty estimates can otherwise be dominated by a Hessian evaluated at a possibly local minimum.

The automation layer retrieves diagnostic data from the MDSplus database \cite{stillerman1997mdsplus}, fits arbitrary time slices for \(n_e\), \(T_e\), \(T_i\), and \(v_T\) in parallel across CPU cores when the corresponding measurements are valid, and feeds the fitted profiles into an MDSplus upload pipeline so that the results become available through the standard KSTAR data access tools.

The paper is organised as follows. Section 2 introduces the modified tanh pedestal profile model and the diagnostic observation models. Section 3 develops the Bayesian formulation, focusing on the robust likelihoods, the default mixture likelihood configuration, and the MCMC sampler. Section 4 illustrates the behaviour of the method on representative cases. It shows robust downweighting of contaminated channels and demonstrates how MAP and posterior mean estimates diverge on a multimodal cost surface. Section 5 describes the software implementation and automated workflow. Section 6 presents operational diagnostics and parallel scaling behaviour relevant to large-scale deployment. Section 7 discusses the significance and limitations of the work, and Section 8 concludes.

\section{Profile and Observation Models}\label{sec:profile-and-forward-models}

\subsection{Modified tanh pedestal profile}\label{sec:modified-tanh-pedestal-profile}

We parametrise each kinetic profile \(f(\psi_N)\) on the normalised poloidal flux coordinate \(\psi_N\) using a modified hyperbolic tangent (mtanh) pedestal profile, following the form introduced for DIII-D pedestal analysis by Groebner et al. \cite{groebner2001mtanh}. The parametrisation is used here as a flexible profile shape, not as the edge pedestal (EPED) physics model itself; however, the fitted pedestal height and width are defined in a way that is compatible with EPED style pedestal workflows \cite{snyder2011eped}. For compact notation, define the edge tanh contribution \(E(\psi_N)\) and the core correction \(C(\psi_N)\) as

\begin{equation}
\begin{aligned}
E(\psi_N)
&= a_1 + a_2\bigl[\tanh(1)-\tanh(z_e)\bigr],\\
C(\psi_N)
&= a_4\Bigl[1-\bigl(z_c\bigr)^{a_5}\Bigr]^{a_6}.
\end{aligned}
\end{equation}

The profile is split at \(\psi_N = 1 - \Delta_{\rm ped}\) into a core region and an edge (pedestal) region:

\begin{equation}
f(\psi_N) =
\begin{cases}
E(\psi_N)+C(\psi_N),
& \psi_N < 1-\Delta_{\rm ped},\\[3pt]
E(\psi_N),
& \psi_N \geq 1-\Delta_{\rm ped},
\end{cases}
\end{equation}
where \(z_e = (\psi_N - 1 + \Delta_{\rm ped}/2)/(\Delta_{\rm ped}/2)\), \(z_c = \psi_N/(1 - \Delta_{\rm ped})\), and the auxiliary constants \(a_2 = (f_{\rm ped} - f_{\rm SOL})/2\) and \(a_1 = [1 - \tanh(1)]\,a_2 + f_{\rm SOL}\), where SOL denotes the scrape-off layer. The core-correction coefficients are \(a_4 = f_{\rm core} - f_{\rm ped}\), \(a_5 = \alpha_{\rm core}\), and \(a_6 = \alpha_{\rm edge}\).

The fitted parameter vector contains seven quantities. Six of them define the mtanh profile shape,

\begin{equation}
\begin{aligned}
\boldsymbol{\theta}_{\rm shape}
&= \bigl(f_{\rm SOL},\; f_{\rm ped}-f_{\rm SOL},\; \Delta_{\rm ped},\\
&\qquad f_{\rm core}-f_{\rm ped},\; \alpha_{\rm core},\; \alpha_{\rm edge}\bigr).
\end{aligned}
\end{equation}
The seventh parameter is the radial shift \(\Delta x\), which absorbs residual mismatches between the diagnostic positions mapped by the equilibrium and the assumed flux surface grid. The sampled vector is therefore \(\boldsymbol{\theta} = \bigl(\boldsymbol{\theta}_{\rm shape},\Delta x\bigr)\). The profile shift is represented as

\begin{equation}
f(\psi_N) \;\longrightarrow\; f(\psi_N - \Delta x).
\end{equation}

The mtanh evaluation is compiled just-in-time so that the profile can be evaluated millions of times per fit at negligible cost; this is essential for the Markov-chain Monte Carlo sampling of Section 3 and for the parallel batch operation of Section 5.

\begin{figure}[tbp]
\centering
\includegraphics[width=0.95\columnwidth]{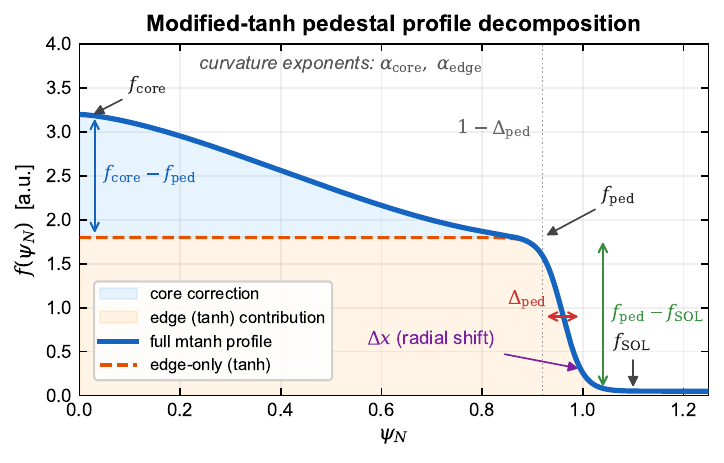}
\caption{Decomposition of the modified tanh pedestal profile into edge tanh and core correction terms. The annotations map each free parameter to a recognizable profile shape change, including pedestal height, SOL value, pedestal width, core curvature, edge curvature, and radial shift.}
\label{fig:mtanh}
\end{figure}

Figure \ref{fig:mtanh} illustrates the decomposition of the mtanh profile into its edge tanh contribution and core correction and annotates every free parameter.

The mtanh model is highly nonlinear, and the associated cost surface is generally non-convex. The pedestal width \(\Delta_{\rm ped}\), the pedestal height \(f_{\rm ped} - f_{\rm SOL}\), and the radial shift \(\Delta x\) enter through the argument and the amplitude of the hyperbolic tangent. This coupling can create multiple local minima when the pedestal is steep and the diagnostic coverage of the edge region is sparse. For a gradient-based optimiser, the choice of initial condition can determine which basin is found. The Hessian error bars computed at a local minimum are systematically too small because they reflect only the local curvature rather than the global landscape. This non-convexity is one of the principal motivations for the MCMC approach developed in Section 3.6.

\subsection{Diagnostic observation models}\label{sec:per-diagnostic-forward-operators}

The workflow does not fit raw detector signals directly. For \(n_e\), \(T_e\), \(T_i\), and \(v_T\), it uses processed diagnostic products retrieved from MDSplus: TS channel values for \(n_e\) and \(T_e\), TCI line integrated density constraints for \(n_e\), and CES derived \(T_i\) and \(v_T\) channel values. For each diagnostic, an observation model maps the fitted one dimensional profile to the corresponding processed diagnostic quantity.

For TS, which supplies \(n_e\) and \(T_e\), the observation model is point evaluation of the candidate profile at the channel positions mapped by the equilibrium, giving \(\hat{\mathbf{y}}^{\rm TS}\). For TCI, the observation model is a chord integral of the candidate \(n_e(\psi_N)\) profile along the five tangential chord paths obtained from the equilibrium, giving the predicted line integrated density \(\hat y^{\rm TCI}_i = \int_{\ell_i} n_e[\psi_N(R(s),Z(s))]\,ds\). For CES, which supplies \(T_i\) and \(v_T\), the observation model is point evaluation at the CES channel positions mapped by the equilibrium. The profile fitter uses processed CES products rather than raw spectra. Conventional beam modulation CES obtains \(T_i\) and \(v_T\) after active/passive spectral separation, whereas the neural-network CES (CESNN) analysis recently developed for KSTAR provides the same processed quantities through an alternative analysis source \cite{lee2026nnces}. The analysis source can be selected through the configuration, and the workflow degrades gracefully when no valid CES channel exists.

The observation models are deliberately modular: each diagnostic owns its own model object exposing a uniform interface for setting the likelihood function. The part specific to each diagnostic is therefore the mapping from a profile to processed diagnostic quantities; the probability densities applied to the resulting residuals are the common Gaussian, Sivia, and mixture forms described in Section 3, with the mixture form used as the default configuration described in Section 3.5.

\section{Bayesian Formulation with Robust Likelihoods}\label{sec:bayesian-formulation-with-robust-likelihoods}

This section develops the Bayesian inference problem by first stating the posterior in its standard form and reviewing the Gaussian baseline. It then introduces two robust alternatives, the conservative likelihood proposed by Sivia and the Gaussian ``good-and-bad'' mixture. The final part explains how the mixture likelihood is configured as the routine default, how the Gaussian and Sivia forms are retained for robustness checks, and how the posterior is sampled with MCMC after a deterministic warm start.

\subsection{Posterior}\label{sec:posterior}

Let \(\mathbf{D} = \{D_k\}_{k=1}^{N}\) denote the concatenated vector of measurements from all enabled diagnostics for a given time slice, let \(\sigma_k\) be the corresponding quoted statistical uncertainty, and let \(\boldsymbol{\theta}\) stand for the seven mtanh parameters of Section 2.1. The observation models of Section 2.2 map \(\boldsymbol{\theta}\) onto a vector of predicted measurements \(\hat{\mathbf{y}}(\boldsymbol{\theta}) = \{\hat y_k(\boldsymbol{\theta})\}\).

Bayes' theorem gives the posterior probability density,

\begin{equation}
p(\boldsymbol{\theta}\mid\mathbf{D}) \;\propto\; p(\mathbf{D}\mid\boldsymbol{\theta})\,p(\boldsymbol{\theta}),
\end{equation}
with \(p(\boldsymbol{\theta})\) a product of independent priors.

A truncated Gaussian is used for parameters with informative prior knowledge, while a uniform prior is used on a physically motivated interval otherwise. For the radial shift parameter \(\Delta x\), the prior regularises a normalised-flux profile-alignment coordinate that absorbs residual equilibrium mapping, diagnostic alignment, and timing mismatches. It should therefore be interpreted as an effective profile shift, not as a literal displacement of diagnostic hardware.

For the default setting used throughout this paper, the bounds on the seven mtanh parameters are \([0,0.1]\), \([0,10]\), \([0.05,0.30]\), \([0,10]\), \([0.8,4]\), \([1,4]\), and \([-0.15,0.05]\) for \(f_{\rm SOL}\), \(f_{\rm ped}-f_{\rm SOL}\), \(\Delta_{\rm ped}\), \(f_{\rm core}-f_{\rm ped}\), \(\alpha_{\rm core}\), \(\alpha_{\rm edge}\), and \(\Delta x\), respectively. The corresponding prior means and standard deviations are chosen on comparable scales within these bounds rather than tuned for a specific time slice, and the numerical settings can be changed by the user when a different operating regime requires it.

Unless stated otherwise, all fits shown in this paper, including the illustrative profile fits, the outlier probability map, and the parallel scaling example, use this same default setting together with the good-and-bad Gaussian mixture likelihood and the settings summarised in Table \ref{tab:mcmc-defaults}.

Conditional on independence between diagnostics, the likelihood factorises,

\begin{equation}
p(\mathbf{D}\mid\boldsymbol{\theta}) \;=\; \prod_{d}\mathcal{L}_d\!\bigl(\mathbf{D}_d\,\big|\,\hat{\mathbf{y}}_d(\boldsymbol{\theta})\bigr),
\end{equation}
and within each diagnostic the channels are taken to be conditionally independent so that \(\mathcal{L}_d\) is itself a product over channels. The functional form of \(\mathcal{L}_d\) is the central modelling choice of this work; the next three subsections describe the Gaussian baseline, the classical Sivia reference form, and the Gaussian mixture used in the deployed workflow. Section 3.5 then explains the operational use of the mixture likelihood and the role of alternatives selected by users.

\subsection{Gaussian baseline}\label{sec:gaussian-baseline}

The classical choice, also the one implicitly assumed by least-squares fitting, is the diagonal Gaussian,

\begin{equation}
\ln \mathcal{L}^{\rm G}_d
= -\,\frac{1}{2}\sum_k \frac{(D_k - \hat y_k)^2}{\sigma_k^2}
- \frac{1}{2}\sum_k \ln(2\pi\sigma_k^2).
\end{equation}
This baseline assumes that every quoted error bar is a faithful description of the true measurement uncertainty. It is optimal, and indeed asymptotically efficient, when this assumption holds, and we retain it as the reference against which the robust alternatives discussed below can be judged.

When the quoted uncertainty is wrong, the quadratic residual penalty gives a single discordant point enough leverage to dominate the likelihood and bias the inferred profile by an arbitrarily large amount, as emphasised by Sivia \cite{sivia1996duff}. The same quadratic cost surface can also sharpen the multimodality of the mtanh model, because a strongly discordant point creates an additional basin into which a deterministic optimiser can fall. The resulting fit can contort the profile to accommodate the outlier at the expense of the bulk of the data. This failure mode is suppressed by the heavy-tailed alternatives described in the next two subsections and by the MCMC exploration of Section 3.6.

\begin{figure}[H]
\centering
\includegraphics[width=0.95\columnwidth]{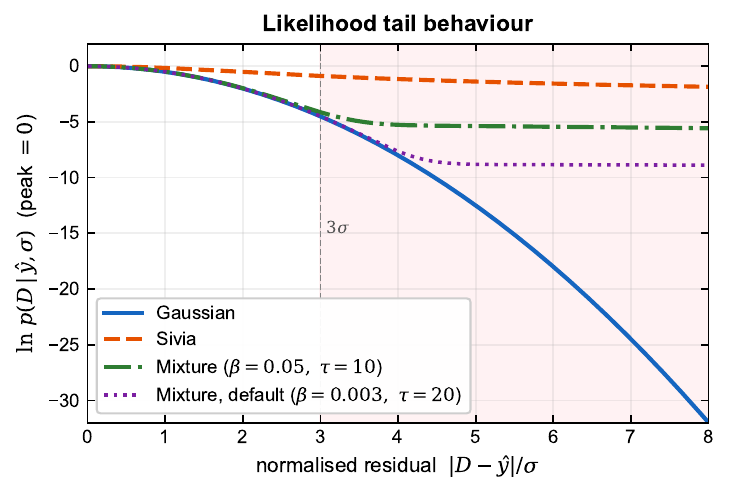}
\caption{Residual-penalty behaviour of the Gaussian, Sivia conservative, and Gaussian-mixture log-likelihoods as a function of normalised residual. The dashed vertical line marks $3\sigma$, and the shaded range highlights larger residuals. The default mixture curve uses $\beta=0.003$ and $\tau=20$, matching the production examples; the second mixture curve illustrates a larger outlier fraction.}
\label{fig:likecompare}
\end{figure}

Figure \ref{fig:likecompare} shows the likelihood forms as a function of the normalised residual \(|D-\hat y|/\sigma\), illustrating the central mechanism by which robust likelihoods prevent outliers from dominating the fit. For residuals of order one, all three likelihoods retain a similar Gaussian-like core and therefore preserve sensitivity to statistically consistent measurements. Beyond roughly \(3\sigma\) (shaded region), however, the Gaussian log-likelihood continues to decrease quadratically, whereas the Sivia conservative likelihood and the Gaussian mixture grow much more slowly over the residual range relevant to the observed outliers. A discordant point therefore receives a much smaller marginal penalty under the robust forms and cannot pull the fitted profile as strongly as it would in a least-squares fit.

\subsection{Conservative robust likelihood (Sivia)}\label{sec:conservative-robust-likelihood-sivia}

For methodological context we also recall the classical conservative formulation of Sivia \cite{sivia1996duff}. It is useful as a reference for the tail-softening mechanism behind robust fitting, but it is not the robust option emphasised in the deployed mixture likelihood workflow of Sections 4-6. We now write \(\sigma_{0k}\) for the quoted uncertainty (denoted \(\sigma_k\) in the preceding sections) to distinguish it from the true, unknown uncertainty for each channel \(\sigma_k\). The quoted error bar \(\sigma_{0k}\) is reinterpreted not as an exact statement of the measurement uncertainty but as a lower bound on it: the true uncertainty \(\sigma_k\) is at least \(\sigma_{0k}\) but may, in principle, be much larger.

This prior encodes the conservative assumption that the quoted uncertainty is a lower bound on the true uncertainty, allowing underestimated error bars to be accommodated without assigning excessive leverage to discordant data. A Jeffreys prior \(p(\sigma_k)\propto 1/\sigma_k\) is placed on the interval \([\sigma_{0k},\sigma_{\max}]\) and the unknown \(\sigma_k\) is marginalised out. Taking the limit \(\sigma_{\max}\to\infty\), the resulting marginal likelihood for each channel is

\begin{equation}
\begin{aligned}
p(D_k\mid\hat y_k,\sigma_k\geq\sigma_{0k})
&\propto \frac{1}{|D_k-\hat y_k|}\\
&\quad \times
\operatorname{erf}\!\left(\frac{|D_k-\hat y_k|}{\sigma_{0k}\sqrt{2}}\right),
\end{aligned}
\end{equation}
so that the full diagnostic log-likelihood is

\begin{equation}
\begin{aligned}
\ln \mathcal{L}^{\rm Sivia}_d
&= \sum_k \ln\!\left[
\frac{1}{|D_k - \hat y_k|}\right.\\
&\quad \left.\times
\operatorname{erf}\!\left(\frac{|D_k - \hat y_k|}{\sigma_{0k}\sqrt{2}}\right)
\right] + \text{const}.
\end{aligned}
\end{equation}
The apparent singularity at \(D_k = \hat y_k\) is removable: \(\operatorname{erf}(x)/x \to 2/\sqrt{\pi}\) as \(x\to 0\), and in the implementation the analytic limit is substituted for numerical stability. The central peak of this distribution has approximately the same width as the original Gaussian, but the tails decay only as \(|D_k-\hat y_k|^{-1}\). A discordant point that would contribute a quadratically diverging penalty under the Gaussian model contributes only a logarithmically growing penalty here, so its leverage on the inferred profile is bounded. The practical advantage of this formulation is that it introduces no free hyperparameters. This simplicity comes at the price of broader posterior uncertainties: even when no outliers are present, the inferred profile parameters can have uncertainties roughly a factor of two larger than those obtained with a Gaussian likelihood, as noted by Sivia \cite{sivia1996duff}.

\subsection{Gaussian ``good-and-bad'' mixture}\label{sec:gaussian-good-and-bad-mixture}

The Sivia formulation is parsimonious but conservative: it increases the posterior width whether or not the data are actually contaminated. For the deployed KSTAR workflow emphasised here, a more efficient model is the ``good-and-bad data'' mixture proposed by Box and Tiao \cite{box1968outlier} and discussed at length by Sivia \cite{sivia1996duff}. When the contamination is rare and well separated from the bulk distribution, as is typically the case for a fringe-jump on a TCI channel or a stray-light spike on a TS chord, this mixture avoids uniform broadening on otherwise clean channels while still suppressing outlier leverage. Each channel is modelled as drawn, with probability \(1-\beta\), from a Gaussian centred on the prediction with the quoted width \(\sigma_{0k}\), and with probability \(\beta\) from a much wider Gaussian with width \(\tau\sigma_{0k}\):

\begin{equation}
\begin{aligned}
N_{0k} &= \mathcal{N}\!\bigl(D_k;\hat y_k,\sigma_{0k}^2\bigr),\\
N_{\tau k} &= \mathcal{N}\!\bigl(D_k;\hat y_k,(\tau\sigma_{0k})^2\bigr),\\
p(D_k\mid\hat y_k,\sigma_{0k},\beta,\tau)
&= (1-\beta)N_{0k}+\beta N_{\tau k},
\end{aligned}
\end{equation}
where \(0\leq\beta\ll 1\) is the outlier fraction and \(\tau\gg 1\) is the outlier width scale. A ``good'' measurement contributes to the first component and behaves exactly as in the Gaussian baseline, while a ``bad'' measurement is absorbed by the broad second component without distorting the fit. The full diagnostic log-likelihood,

\begin{equation}
\ln \mathcal{L}^{\rm mix}_d
= \sum_k \ln\!\Bigl[(1-\beta)N_{0k}+\beta N_{\tau k}\Bigr],
\end{equation}
is evaluated channel-by-channel using a numerically stable log-sum-exp so that the small ``good'' component does not underflow when the residual is large.

The mixture likelihood smoothly interpolates between the Gaussian baseline (recovered as \(\beta\to 0\) or \(\tau\to 1\)) and a heavy-tailed distribution (large \(\beta\), large \(\tau\)). The deployed workflow uses fixed, configurable values for these mixture parameters because this is stable for unattended production runs and matches the settings exposed to users. The production examples in this paper use the default fixed values \(\beta=0.003\) and \(\tau=20\), while the user-facing interfaces allow these settings to be changed when needed.

The same numerical settings are applied consistently across fitted quantities by expressing profile amplitudes in fixed engineering units: \(n_e\) in \(10^{19}\,\mathrm{m}^{-3}\), \(T_e\) and \(T_i\) in \(\mathrm{keV}\), and \(v_T\) in \(100\,\mathrm{km\,s}^{-1}\). These unit choices put the fitted amplitudes and their quoted uncertainties on comparable order-of-magnitude scales, so that shared robustness parameters can be used without retuning their numerical interpretation for each quantity.

The numerical meaning of these defaults is best understood relative to the uncertainty scale of the fitted diagnostic data. As a representative check, for KSTAR shot 32364 over \(2.0\)--\(8.0\,\mathrm{s}\) with \(\Delta t=0.05\,\mathrm{s}\), the median quoted uncertainties are \(0.466\times 10^{19}\,\mathrm{m}^{-3}\) for TS \(n_e\), \(0.214\,\mathrm{keV}\) for TS \(T_e\), \(0.080\,\mathrm{keV}\) for \(T_i\) from the CESNN analysis of Ref.~\cite{lee2026nnces}, and \(3.84\,\mathrm{km\,s}^{-1}\) for CESNN \(v_T\). Expressed in the engineering units above, the typical single-channel uncertainty is of order \(10^{-2}\)--\(10^{-1}\) for the rotation profile, order \(10^{-1}\) for temperature profiles, and order \(10^{-1}\)--\(1\) for density.

With \(\tau=20\), the broad component is therefore much wider than ordinary channel scatter. Using the representative median uncertainties, its standard deviation is approximately \(9.3\) in \(10^{19}\,\mathrm{m}^{-3}\) for \(n_e\), \(4.3\,\mathrm{keV}\) for \(T_e\), \(1.6\,\mathrm{keV}\) for \(T_i\), and \(0.77\) in \(100\,\mathrm{km\,s}^{-1}\) units for \(v_T\). The small value \(\beta=0.003\), however, keeps this broad component rare. For \(\beta=0.003\) and \(\tau=20\), the posterior probability of belonging to the broad component exceeds \(50\%\) only at a normalised residual of about \(|D_k-\hat y_k|/\sigma_{0k}\simeq 4.2\). Therefore ordinary \(2\)--\(3\sigma\) scatter remains effectively Gaussian, while isolated multi-\(\sigma\) excursions are absorbed by the broad component and downweighted in the fit. These values should be interpreted as operational robustness settings rather than as universal physical probabilities of diagnostic failure.

This mixture treatment yields a useful diagnostic quantity. Given a posterior sample \(\boldsymbol{\theta}^{(s)}\) and the corresponding model prediction \(\hat y_k^{(s)}\), define \(\beta\) and \(\tau\) as the configured mixture parameters. The posterior probability that the \(k\)-th channel is an outlier is then, by Bayes' rule applied to the mixture indicator,

\begin{equation}
P^{(s)}_k =
\frac{\beta N^{(s)}_{\tau k}}
{(1-\beta)N^{(s)}_{0k}+\beta N^{(s)}_{\tau k}}.
\end{equation}
Here \(N^{(s)}_{0k}\) and \(N^{(s)}_{\tau k}\) are evaluated with the sampled prediction and the configured values of \(\beta\) and \(\tau\). Averaging \(P^{(s)}_k\) over the chain produces a scalar for each channel in \([0,1]\) that we report alongside the fitted profile as an automatic flag of which measurements have been effectively downweighted. This quantity is useful for downstream data quality monitoring: candidate systematics in individual channels or underestimated uncertainties appear as a persistent population of high probability outliers across many time slices, in a way that is not accessible to a Gaussian fit followed by manual cleaning.

\subsection{Operational mixture likelihood configuration}\label{sec:operational-mixture-likelihood-configuration}

For operational use with several diagnostics, the failure state is not known in advance. Different diagnostics have qualitatively different failure modes. TS edge channels are vulnerable to stray light and to plasma--wall interaction artefacts. TCI chords can suffer from fringe jumps that produce isolated \(5\sigma\)--\(10\sigma\) excursions. CES channels close to the edge can return non-physical values when the carbon emission becomes too weak. The operational strategy used here is therefore to apply the good-and-bad Gaussian mixture as the default likelihood for fitted diagnostic channels. When the data are consistent, the posterior weight remains on the narrow Gaussian component and the fit behaves close to the Gaussian baseline; when an individual measurement is discordant, the broad component limits its leverage and produces a quantitative outlier probability.

The standard Gaussian and the Sivia conservative likelihood are retained as alternatives selectable by users for sensitivity studies, but the default batch and GUI workflows use the mixture likelihood unless the user explicitly overrides it.

\subsection{MCMC sampler}\label{sec:mcmc-sampler}

A deterministic optimiser, whether gradient-based (quasi-Newton, Levenberg--Marquardt) or derivative free (Nelder--Mead \cite{nelder1965simplex}), returns a single point estimate and, at best, a Hessian derived covariance matrix evaluated at that point. For a model that is globally convex this is entirely adequate. The mtanh profile, however, is generally non-convex in the joint space of pedestal width \(\Delta_{\rm ped}\), pedestal height \(f_{\rm ped}\), and radial shift \(\Delta x\): different combinations of these three parameters can produce qualitatively different profile shapes that fit the sparse edge data comparably well, giving rise to multiple local minima separated by narrow ridges in the cost surface.

The phenomenon is especially pronounced for steep H-mode pedestals, where a centimetre-scale effective radial displacement associated with \(\Delta x\) near the pedestal can interchange the roles of the pedestal foot and the SOL shoulder, creating an alternative basin with a cost that is only marginally worse than the global minimum yet corresponds to a qualitatively different physical interpretation.

A purely optimisation based approach offers no reliable way to distinguish the global minimum from such a secondary basin. Restarting the optimiser with multiple initial conditions can help, but the number of restarts needed grows with the dimensionality and the complexity of the cost surface, and there is no guarantee that the global optimum has been found. Crucially, the Hessian derived error bars evaluated at a local minimum are conditional on that minimum being the global one, an assumption that is false precisely when it matters most. The resulting error bars can be overly narrow, and this underestimated uncertainty can propagate into downstream analyses.

MCMC sampling mitigates this problem. Given sufficient run length, the chain can explore multiple basins of the posterior and, provided that inter-mode mixing is adequate, the posterior mean and credible intervals integrate over the accessible landscape rather than conditioning on a single mode. Even if the chain is initialised from a MAP estimate that happens to sit in a secondary basin, stochastic transitions can carry it into higher-probability regions; if a secondary mode retains appreciable posterior mass and is reachable on the sampled time scale, the resulting posterior uncertainty reflects the multimodal structure more faithfully than any single-point estimate.

In the present approach, a bounded L-BFGS-B minimisation of the negative log posterior is first used to obtain a MAP-seeking point estimate. This point is used only as a warm start for the sampler, not as a final uncertainty estimate. Walkers are initialised by perturbing this point within a small neighbourhood and clipping the perturbed values to the prior bounds, so that the ensemble starts in a high density region while the burn-in phase relaxes the initial condition.

The posterior is then explored with the emcee implementation of the affine-invariant ensemble sampler \cite{goodman2010ensemble,foremanmackey2013emcee}. The sampler deploys \(N_{\rm walk}=28\) walkers for the default seven-parameter mtanh fit. The default workflow runs \(5000\) ensemble steps and discards the first \(4000\) ensemble steps as burn-in; retained walker samples may be thinned or capped in the output files for storage.

The sampler settings used by the default workflow are summarised in Table \ref{tab:mcmc-defaults}. The combination of a MAP-seeking warm start, likelihood evaluations compiled just-in-time with Numba, and the emcee ensemble sampler has been practical for unattended batch processing with a fixed configuration. Section 4.2 illustrates how the MAP and posterior mean estimates diverge on a concrete multimodal case. One ensemble step advances all walkers once, so \(N_{\rm step}\) ensemble steps produce \(N_{\rm step}N_{\rm walk}\) raw walker samples before burn-in removal. These are correlated MCMC samples rather than independent effective samples; the effective sample size depends on the autocorrelation time of each parameter.

\begin{table}[tbp]
\centering
\small
\begin{ruledtabular}
\begin{tabular}{l|l}
\makebox[0.36\columnwidth][l]{Setting} & \parbox[t]{0.54\columnwidth}{Default value} \\
\hline
\makebox[0.36\columnwidth][l]{Sampler} & \parbox[t]{0.54\columnwidth}{emcee affine-invariant} \\
\makebox[0.36\columnwidth][l]{Dimension} & \parbox[t]{0.54\columnwidth}{\(d=7\)} \\
\makebox[0.36\columnwidth][l]{Walkers} & \parbox[t]{0.54\columnwidth}{\(N_{\rm walk}=4d=28\)} \\
\makebox[0.36\columnwidth][l]{Steps} & \parbox[t]{0.54\columnwidth}{\(N_{\rm step}=5000\)} \\
\makebox[0.36\columnwidth][l]{Burn-in} & \parbox[t]{0.54\columnwidth}{\(4000\) steps} \\
\makebox[0.36\columnwidth][l]{Retained} & \parbox[t]{0.54\columnwidth}{\(1000N_{\rm walk}=2.8\times10^4\)} \\
\makebox[0.36\columnwidth][l]{Mixture fraction} & \parbox[t]{0.54\columnwidth}{\(\beta=0.003\)} \\
\makebox[0.36\columnwidth][l]{Mixture scale} & \parbox[t]{0.54\columnwidth}{\(\tau=20\)} \\
\end{tabular}
\end{ruledtabular}
\caption{Default MCMC and mixture likelihood settings. Retained samples
are raw correlated walker samples before optional thinning or capping.}
\label{tab:mcmc-defaults}
\end{table}

\section{Illustrative Behaviour of the Method}\label{sec:illustrative-behavior-of-the-method}

This section uses two representative KSTAR H-mode slices to separate the two key mechanisms of the method: robust downweighting of contaminated channels and posterior exploration on a multimodal surface. To keep these mechanisms visually distinct, the illustrative examples below use the same fixed mixture parameters as the default workflow; Section 6 then reports operational diagnostics and scaling behaviour relevant to large-scale deployment.

\subsection{Gaussian vs mixture likelihood on contaminated data}\label{sec:gaussian-vs-mixture-likelihood-on-contaminated-data}

\begin{figure}[tbp]
\centering
\includegraphics[width=0.95\columnwidth]{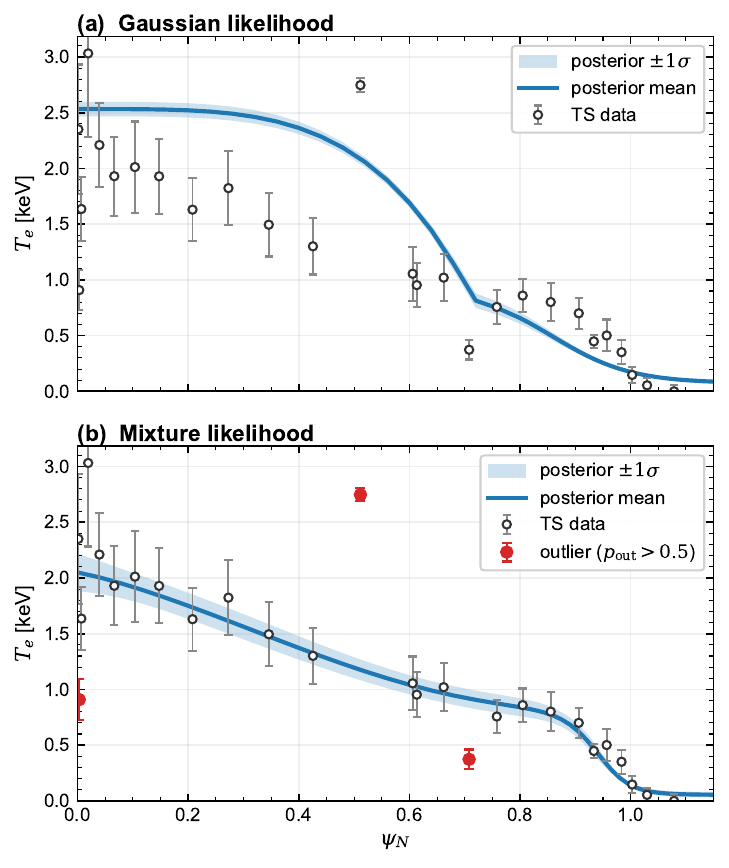}
\caption{Gaussian and Gaussian-mixture fits for TS-$T_e$ in KSTAR shot 32364 at $t=4.79\,\mathrm{s}$ using identical data and priors. The Gaussian fit is pulled by discordant channels and broadens the pedestal, whereas the mixture fit flags those channels as outliers and preserves the steep edge.}
\label{fig:gaussmix}
\end{figure}

Figure \ref{fig:gaussmix} illustrates the effect of the likelihood choice on a TS electron temperature time slice from KSTAR shot 32364 at \(t=4.79\,\)s, an H-mode discharge in which a small number of TS channels returned values that depart visibly from the bulk profile. Both fits use identical data and priors; the only difference is the likelihood applied to the TS-\(T_e\) channel group.

Panel (a) shows the result of the standard Gaussian likelihood. The three discordant channels, located near the magnetic axis (\(\psi_N \approx 0\)), at mid-radius (\(\psi_N \approx 0.5\)), and near the pedestal foot (\(\psi_N \approx 0.7\)), receive the same statistical weight as the rest of the data. The optimiser is forced to compromise. The inferred pedestal width inflates to \(\Delta_{\rm ped} = 0.283\), more than twice the mixture-fit value shown in panel (b). The core curvature is pushed to the prior boundary in an attempt to pull the profile upward toward the mid-radius outlier, and the resulting profile has a broad, shallow pedestal that is inconsistent with the steep H-mode edge inferred from the bulk of the measurements.

The core temperature is overestimated by approximately \(0.5\,\)keV. Such a distortion is problematic in unattended operation because the Gaussian fit can converge normally and report tight error bars while still producing a pedestal gradient that is unsuitable for downstream equilibrium or stability analysis.

Panel (b) shows the same data fitted with the Gaussian ``good-and-bad'' mixture likelihood (Section 3.4). The mixture parameters are fixed at the default values \(\beta = 0.003\) and \(\tau = 20\) so that the visual difference relative to the Gaussian fit reflects only the effect of the robust likelihood shape. Of the \(27\) enabled channels, three are automatically assigned a posterior outlier probability greater than \(50\%\) and are highlighted in red. The posterior mean profile and its \(\pm 1\sigma\) band track the bulk of the channels and preserve a sharp pedestal with \(\Delta_{\rm ped} = 0.121\) and a smooth core slope; no manual masking was applied. The pedestal top (\(0.83\,\mathrm{keV}\)) and core temperature (\(2.05\,\mathrm{keV}\)) are both consistent with the channels assigned low outlier probability, and the outlier probabilities for each channel provide an automatic record of which measurements were downweighted.

\subsection{MAP and posterior mean under the default broad prior}\label{sec:map-vs-posterior-mean-on-a-multimodal-cost-surface}

\begin{figure}[tbp]
\centering
\includegraphics[width=0.95\columnwidth]{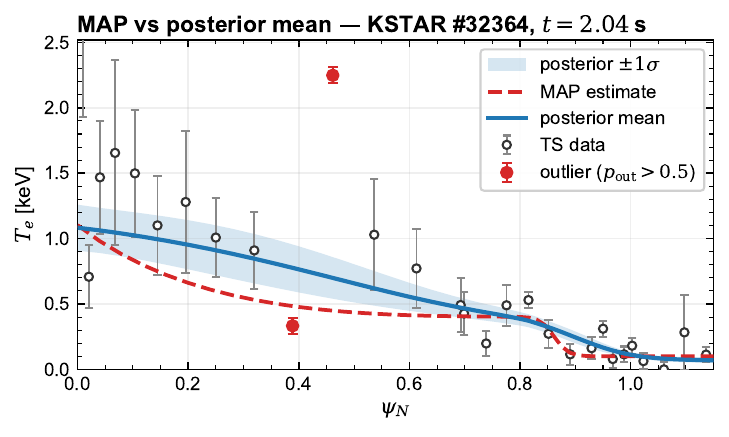}
\caption{MAP estimate and posterior mean profile for the $T_e$ fit of KSTAR shot 32364 at $t=2.04\,\mathrm{s}$ using the default broad prior. The MAP curve is pulled toward a narrow pedestal, strongly shifted solution, while the posterior mean integrates over the broader sampled posterior.}
\label{fig:map-post}
\end{figure}

Figure \ref{fig:map-post} shows a \(T_e\) fit for KSTAR shot 32364 at \(t=2.04\,\)s under the same broad default prior listed in Section 3.1. The mixture parameters are fixed at \(\beta=0.003\) and \(\tau=20\), so that the displayed posterior structure reflects the mtanh parameter geometry rather than additional uncertainty in the mixture parameters.

Because this default prior includes a broad radial shift interval, the posterior has more freedom to trade pedestal position against pedestal width and core shape. In this example the MAP estimate (red dashed) and the posterior mean (blue solid) separate visibly in the core region while remaining similar near the pedestal top and SOL. The parameter summaries differ even more strongly: the MAP estimate is located on the lower pedestal-width boundary, with \(\Delta_{\rm ped}=0.050\), and uses a large negative normalised-flux shift, \(\Delta x=-0.114\). This MAP value should be read as a profile-alignment nuisance parameter within the deliberately broad default interval, not as a direct estimate of the equilibrium error.

The posterior mean, by contrast, gives \(\Delta_{\rm ped}=0.195\) and \(\Delta x=-0.004\). A single optimiser output would therefore misrepresent both the profile-level uncertainty and the correlated parameter uncertainty.

\begin{figure*}[t!]
\centering
\includegraphics[width=0.94\textwidth,height=0.74\textheight,keepaspectratio]{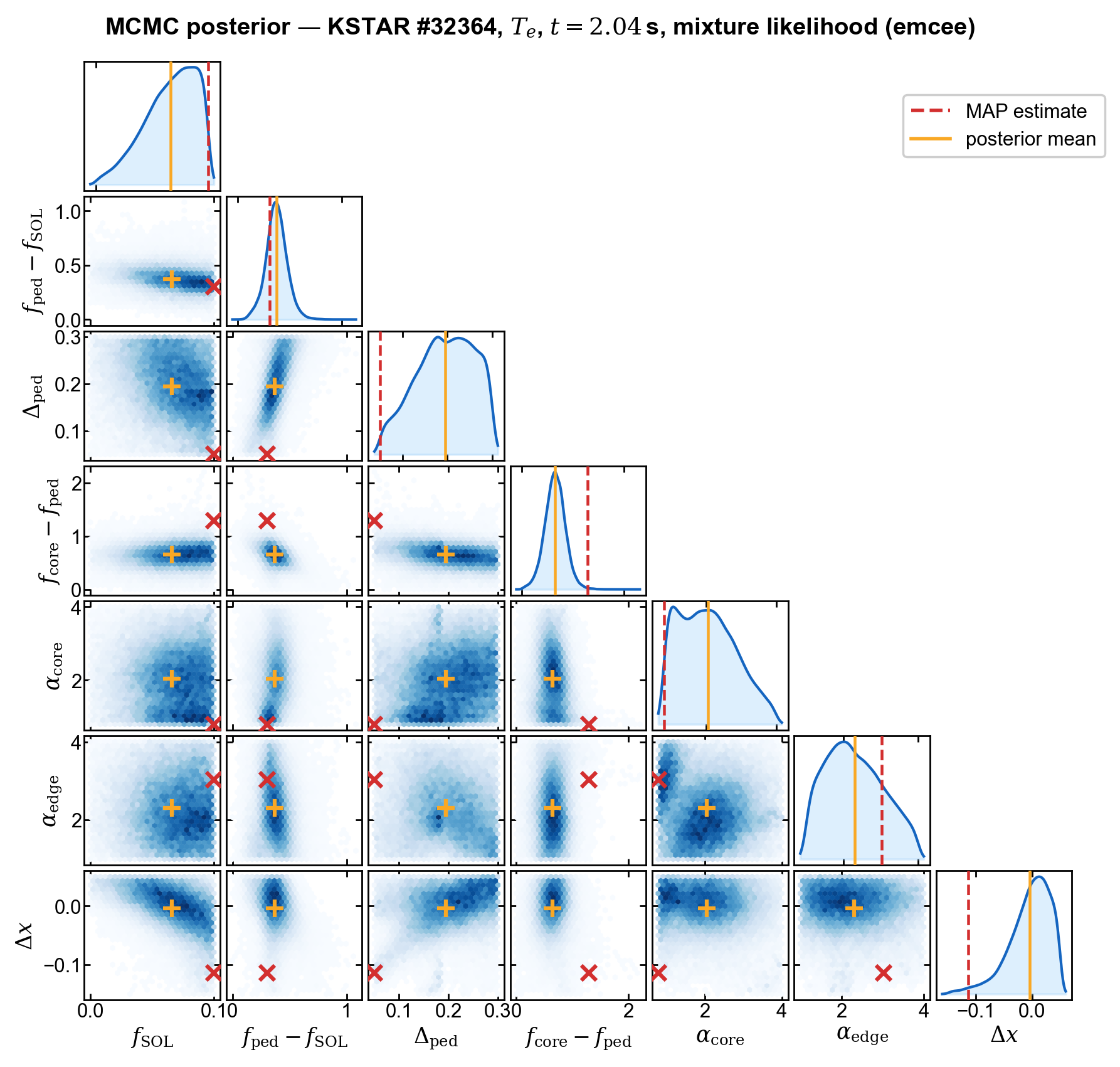}
\caption{Corner plot of the MCMC posterior for the $T_e$ fit shown in Figure \ref{fig:map-post}. All seven mtanh parameters are shown, including the radial shift $\Delta x$. The broad prior exposes correlations among pedestal width, SOL value, radial shift, and core-shape parameters that are not visible from the MAP profile alone. The $\alpha_{\rm core}$ marginal distribution and the $\alpha_{\rm core}$-dependent joint distributions with $\alpha_{\rm edge}$, $f_{\rm SOL}$, and $f_{\rm ped}-f_{\rm SOL}$ show two separated lobes. This structure is consistent with a multimodal posterior rather than a single skewed or elongated basin.}
\label{fig:corner}
\end{figure*}

The corresponding corner plot in Figure \ref{fig:corner} includes all seven mtanh parameters, including the radial shift \(\Delta x\). To visualise the posterior structure clearly, this diagnostic corner plot uses a longer MCMC chain, retaining \(1.87\times 10^5\) raw walker samples, rather than the default batch setting in Table \ref{tab:mcmc-defaults}. It shows why profile-level agreement between MAP and posterior mean is not equivalent to parameter-level certainty. The broad prior allows \(\Delta x\) to move over a wider interval, and the posterior uses this freedom: negative radial shifts are correlated with larger pedestal widths and different SOL values. The MAP point lies near the edge of the sampled support in several coordinates, particularly \(\Delta_{\rm ped}\), \(f_{\rm core}-f_{\rm ped}\), and \(\Delta x\), where the displacement from the posterior mean reaches several posterior standard deviations. The posterior mean should therefore be interpreted as a posterior-averaged summary of the profile function, not as a claim that the mean parameter vector defines a unique physical mode.

The most visible multimodality appears in the core-shape parameters. The \(\alpha_{\rm core}\) posterior separates into two branches that are correlated with \(\alpha_{\rm edge}\), \(f_{\rm SOL}\), and \(f_{\rm ped}-f_{\rm SOL}\). These branches correspond to different ways of distributing curvature between the core correction and the edge tanh part of the mtanh profile while producing similar profile values over the measured radial interval. A deterministic optimiser would select one branch and report uncertainties conditional on that branch, whereas the MCMC samples expose both branches in the posterior summary.

This example illustrates the role of MCMC in the automated workflow. A deterministic optimiser can provide a useful starting point and often a reasonable profile curve, but it cannot by itself report the correlated uncertainty caused by the broad prior and the non-convex mtanh geometry. The sampled posterior exposes those correlations directly and allows downstream users to distinguish a well-constrained profile function from poorly constrained individual shape parameters.

\section{Software Implementation and Automated Workflow}\label{sec:software-implementation-and-automated-workflow}

The mathematical formulation of Sections 3--4 is realised as a software pipeline with explicit separation between data access, diagnostic observation modelling, single time slice fitting, batch orchestration, output serialisation, and MDSplus upload. Diagnostic and equilibrium data are retrieved from the MDSplus database \cite{stillerman1997mdsplus} before the observation-model layer constructs the diagnostic mappings described in Section 2.2. The per-time-slice fit module then runs the MAP optimisation and MCMC sampler of Section 3, after which the orchestration layer distributes independent slices across CPU cores. Results are serialised to JSON or HDF5 and can be published to a dedicated MDSplus tree for downstream consumers. An interactive graphical frontend and a noninteractive batch frontend share the same fit module, so that an interactive fit and a batch fit produce numerically identical results under the same configuration and random seed.

The implementation is written entirely in Python 3.10+. Numerical work uses NumPy and SciPy for array operations and the quasi-Newton optimiser used for the MAP estimate. The inner loops of the profile evaluation and likelihood functions are compiled just-in-time with Numba, yielding per-evaluation costs comparable to a compiled language. Posterior sampling uses the affine-invariant ensemble sampler in emcee \cite{goodman2010ensemble,foremanmackey2013emcee}, while the standard library's multiprocessing module provides cross-process parallelism. Output is written through JSON and HDF5 writers, diagnostic access is provided by MDSplus, and the graphical interface is built with PyQt5 (Qt 5). The package requires no custom compiled extension modules beyond the standard scientific Python stack and MDSplus, making it suitable for deployment on a shared analysis server.

\subsection{Per-time-slice fit module}\label{sec:per-time-slice-fit-module}

The atomic unit of work is the fit of a single physical quantity at a single time slice. Given the equilibrium reconstruction at the requested time, the diagnostic data and error bars, the channel-use masks, the prior and initial-value configuration, and the likelihood configuration of Section 3.5, the fit module constructs the diagnostic forward operators of Section 2.2 and attaches the requested likelihood. It then builds the joint posterior, computes the MAP estimate, runs the MCMC chain of Section 3.6, and post-processes the samples to produce the MAP profile, the posterior mean and standard deviation, and the outlier probabilities produced by the mixture likelihood for each channel. The fit module is deliberately stateless and consumes only plain data, which is what allows it to be dispatched across worker processes without serialisation issues.

\subsection{Batch operation and parallelisation}\label{sec:batch-operation-and-parallelism}

The noninteractive frontend takes a shot number, an equilibrium tree, a list of physical quantities to be fitted, the likelihood configuration, and either a uniform \(\Delta t\) time grid or an explicit list of time slices, and processes the entire shot in a single command. All diagnostic data are retrieved from MDSplus exactly once at the start of the run and then held in memory while the per-time-slice fits are dispatched to a configurable pool of worker processes.

Several auxiliary controls support campaign-database operation. Time intervals can be filtered by a plasma-current threshold, with a configurable margin from the first and last crossings so that the breakdown and ramp-down phases are excluded. Individual diagnostic channels can be permanently masked for shots where they are known to be unreliable, and the MDSplus cache can be reused across runs to avoid redundant data access. A wrapper script chains the batch fit with the downstream upload step so that a single command produces fitted profiles, writes them to local storage, and publishes them to the MDSplus tree consumed by downstream codes.

In large shot lists, some requested time intervals may lack one or more processed diagnostic products because of startup or wall conditioning phases, unavailable beam modulation CES products, or diagnostic maintenance and data acquisition issues. The package is engineered so that a missing diagnostic for one quantity does not abort the run for the others: each loader either returns valid data or raises a typed exception that the orchestration layer catches, converts into a structured warning, and uses to skip only the affected quantity while the remaining quantities continue to be fitted. This level of fault tolerance is essential for unattended batch operation over a campaign database.

A Qt graphical interface (PyQt5) that shares the same fit module and exposes the same advanced likelihood options and numerical configuration controls as the batch frontend is described separately in Appendix A.

\subsection{Output and downstream pipeline}\label{sec:output-and-downstream-pipeline}

For each fitted time slice the package writes a self-contained file in either JSON or HDF5 format. JSON is used when human readability and convenient inspection are useful, whereas HDF5 is used for batch-database operation. The file stores the MAP profile, the posterior mean and standard deviation, a configurable maximum number of posterior samples, the channel use masks, the outlier probabilities for each channel where applicable, the likelihood configuration, and the random seed used to drive the MCMC chain. A companion upload tool reads these files and publishes the same data into a dedicated MDSplus tree, so that researchers without local access to the fit output can consume the results through the standard MDSplus interfaces. Once uploaded, the fitted profiles, posterior uncertainties, and channel-wise outlier probabilities can also be inspected through PRISM, the integrated visualisation platform used in KSTAR analysis \cite{lee2026prism}.

\section{Operational Diagnostics and Scaling Toward Large Data Sets}\label{sec:campaign-scale-deployment-and-operational-diagnostics}

Whereas Section 4 used single time slice illustrative examples to isolate specific mechanisms, this section focuses on operational outputs needed for future large-scale deployment: outlier monitoring products, batch execution behaviour, and parallel scaling under the default workflow.

\subsection{Batch processing and outlier monitoring}\label{sec:database-processing-and-outlier-monitoring}

With the default mixture likelihood configuration described in Section \ref{sec:operational-mixture-likelihood-configuration}, the package can process arbitrary shot/time lists and write fitted profiles in JSON/HDF5 form for MDSplus upload and downstream analysis workflows. This batch mode is intended to support upcoming large-scale KSTAR kinetic profile production for kinetic-EFIT reconstruction, TRANSP integrated modelling, and machine learning data set construction. In that setting, the fitted profiles provide not only profile values but also uncertainty estimates, fit metadata, and data quality indicators that are needed to curate profile data sets without visually inspecting every time slice.

\begin{figure*}[t]
\centering
\includegraphics[width=0.94\textwidth]{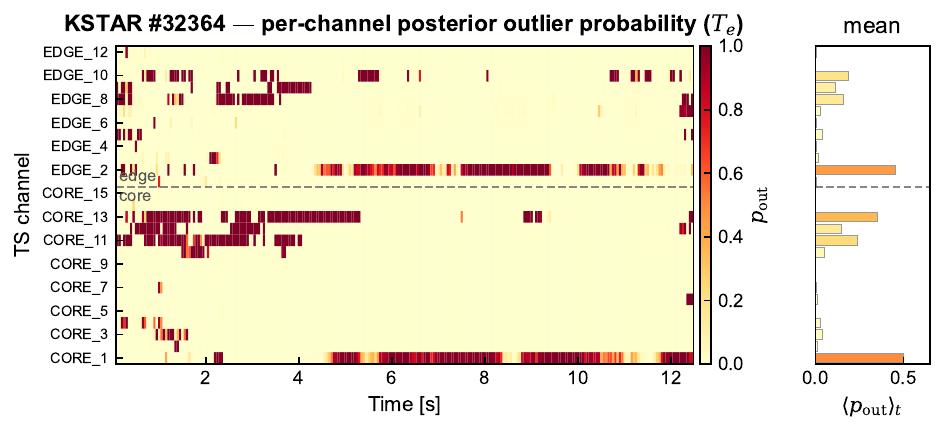}
\caption{Posterior outlier probability map for TS-$T_e$ across 248 fitted time slices of KSTAR shot 32364. The shortened vertical range covers CORE\_1 through EDGE\_12 using the one-based MDSplus/diagnostic channel convention. The map is intended as a screening tool for repeated disagreement between the fitted model and the data and for follow-up inspection in large-scale fitting campaigns.}
\label{fig:outlier-map}
\end{figure*}

A valuable by-product of the batch processing is the outlier probability map for each channel. Figure \ref{fig:outlier-map} shows the posterior outlier probability for the displayed TS channel range from CORE\_1 through EDGE\_12 across every fitted time slice of KSTAR shot 32364 (\(T_e\), 248 time slices, mixture likelihood). The channel labels follow the one-based MDSplus/diagnostic convention used operationally; every other label is printed to keep the shortened vertical axis readable.

The heatmap reveals two distinct patterns. CORE\_1 remains strongly downweighted over much of the discharge, while CORE\_11, CORE\_13, EDGE\_2, and EDGE\_10 show elevated time-averaged outlier probabilities in this shot.

The map provides an automated and reproducible summary of where the statistical model found repeated tension between individual measurements and the fitted profile. Such tension can arise from several sources. Beyond instrumental or analysis effects, real plasma behaviour can also produce outlier signatures: time intervals containing internal transport barriers, fast pedestal transients, or other profile structures not represented by a single mtanh function can be flagged because they lie outside the model space rather than because the measurements themselves are faulty. The outlier probability map should therefore be interpreted as showing where the fitted model and the data are mutually inconsistent within the assumed uncertainties, requiring contextual interpretation, rather than as a direct classifier of diagnostic faults.

This kind of diagnostic quality monitoring comes at no additional computational cost because the outlier probabilities are already computed as part of the standard fit, and it provides a practical guide for collaborative follow-up with diagnostic experts by helping to prioritise channels or time intervals for inspection in large-scale fitting campaigns.

\subsection{Computational scaling}\label{sec:computational-scaling}

\begin{figure}[tbp]
\centering
\includegraphics[width=0.95\columnwidth]{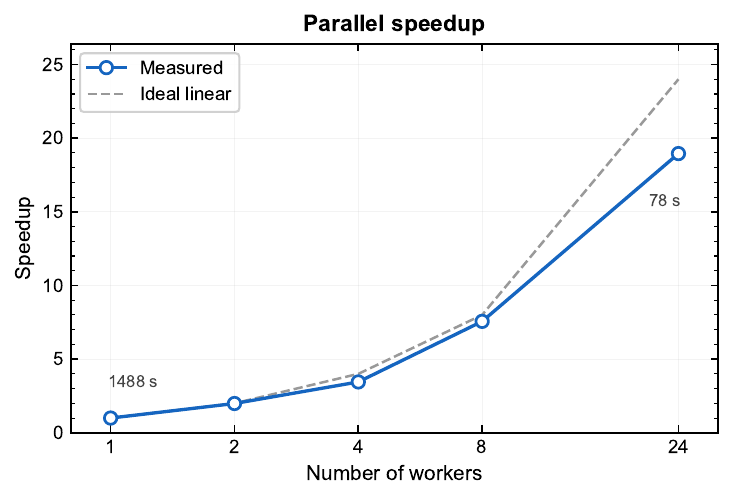}
\caption{Parallel speedup of pure fitting time for KSTAR shot 32364 ($T_e$, 248 time slices, 5000 ensemble steps with 4000 burn-in steps), excluding the one-time data loading and job construction phase. The displayed runtimes are normalised from the measured per-slice timing to the same 248-slice workload used in Figure \ref{fig:outlier-map}. The benchmark was measured on AMD EPYC 7453 processors. Independent posterior evaluation for each slice gives near-linear speedup up to eight workers and remains useful at 24 workers.}
\label{fig:wallclock}
\end{figure}

Figure \ref{fig:wallclock} quantifies the parallel speedup on pure fitting time (MDSplus data retrieval excluded) for a representative KSTAR shot (32364, 248 \(T_e\) time slices) using the default sampling configuration of \(5000\) ensemble steps with \(4000\) burn-in steps. The benchmark was measured on AMD EPYC 7453 processors, and the displayed runtimes are normalised to the same 248-slice workload used in Figure \ref{fig:outlier-map} by multiplying the measured per-slice timing. The single-worker run takes \(1488\,\)s, the eight-worker run takes \(197\,\)s, and the 24-worker run takes \(78\,\)s, corresponding to measured speedups of \(7.6\times\) and \(19.0\times\), respectively.

The nearly linear behaviour up to eight workers shows that independent posterior evaluation for each slice parallelises efficiently in the regime normally used for interactive batch operation. The 24-worker run reduces the pure fitting time for 248 \(T_e\) slices to about \(78\,\)s, which is sufficiently fast for practical automated processing. In KSTAR operation, the interval between plasma shots is typically on the order of ten minutes; the measured runtime therefore indicates that fitted profiles, uncertainty summaries, metadata, and profile plot PDF outputs can be generated and inspected within the interval between shots for representative cases.

The serial data loading and job construction phase is excluded from the timing and runs only once regardless of the number of workers; in this benchmark that phase took less than \(1\,\)s. The result should therefore be read as the scaling of the posterior evaluation and fitting layer rather than as an end-to-end MDSplus data access benchmark.

\section{Discussion}\label{sec:discussion}

\subsection{Significance}\label{sec:significance}

The work described here addresses a workflow gap in the KSTAR analysis chain. Before this work, profile fitting was carried out interactively, one shot at a time, by individual researchers with their own scripts and their own subjective channel masks. This made it impractical to assemble large kinetic profile data sets and, as a consequence, made it difficult to prepare reproducible inputs for systematic transport, pedestal, confinement scaling, kinetic-EFIT, or TRANSP studies beyond a small number of selected discharges.

The present pipeline turns profile fitting into a routine production step: a single command processes an entire shot, a wrapper command processes shot/time lists, and the MDSplus upload pipeline makes the results available through the standard KSTAR data access interfaces.

Another potential application is the construction of labelled data sets for machine learning algorithms that detect outliers in KSTAR diagnostic measurements. The mixture likelihood used here produces posterior outlier probabilities for measurements at each channel and time slice, which can serve as soft labels or as an initial candidate labelling tool for supervised or weakly supervised outlier classifiers. These labels should not be interpreted as ground truth diagnostic faults, because they measure tension between the fitted profile model and the data relative to the quoted uncertainty, equilibrium mapping, and mtanh profile model. Nevertheless, they provide a reproducible and scalable starting point for building curated training data sets with subsequent review by diagnostic experts.

The deployment on the shared KSTAR analysis server, with multiuser safe cache directories for each user and read/access permissions for all users, was a deliberate engineering choice. This deployment choice reflects the practical requirement that an operational analysis tool must be runnable by users other than its original developer.

From a scientific perspective, the work demonstrates that likelihoods tolerant to outliers and MCMC sampling, applied together, mitigate the principal failure modes encountered in automated profile fitting on a real tokamak. It also shows that the resulting pipeline is practical as an operational profile generation layer for downstream equilibrium and transport analysis.

One implication concerns outlier robustness. The robust likelihood ideas of \cite{sivia1996duff,box1968outlier}, originally discussed in the contexts of laboratory mean estimation and general Bayesian outlier modelling, can be adapted to a tokamak profile fit with multiple diagnostics. In the deployed KSTAR workflow, the key practical step is the use of the good-and-bad Gaussian mixture as the default likelihood for fitted diagnostic channels. This gives a common robust treatment across the workflow while allowing clean measurements to remain effectively Gaussian through the narrow component with high probability.

The classical Sivia conservative form remains useful as a selectable robust reference for the same tail-softening mechanism, while the mixture model is the likelihood used in routine production and in the examples shown here. The outlier probabilities for each channel provide a quantitative measure of data quality on a slice by slice basis and can be used to identify candidate systematics in individual channels or time intervals that deserve closer diagnostic review.

A second implication concerns local minimum robustness. Equally important is the recognition that the mtanh cost surface is generically non-convex and that the standard practice of reporting a single optimiser output as ``the'' fit is inadequate for automated batch operation. A deterministic optimiser may converge to a secondary basin with a pedestal width or radial shift that is not representative of the sampled posterior, and the resulting Hessian error bars will be too small because they are conditional on the wrong mode.

The MCMC chain, by contrast, can reveal multimodal posterior structure when mixing between modes is adequate, and the posterior mean and credible intervals then integrate over the sampled contributions. Even when the posterior is effectively unimodal, the MCMC error bars capture nonlinear curvature more directly than a local quadratic approximation. This property is useful for downstream consumers of the fitted profiles, including kinetic-EFIT and integrated modelling codes such as TRANSP \cite{pankin2025transp}, whose inferred quantities are sensitive to local profile gradients.

The same strategy enables downstream automation. The combination of automatic outlier handling and posterior exploration reduces the need for visual inspection of every time slice by providing reproducible posterior summaries and data quality diagnostics. The MDSplus upload pipeline described in Section 5 is designed so that fitted profiles can be consumed by kinetic-EFIT and batch transport codes without manual reformatting. The long term goal is a pipeline from shot to analysis in which kinetic profile generation becomes a standard automated input stage for kinetic-EFIT, TRANSP, and data-driven profile analyses, while the human role is focused on monitoring outlier probability maps for each channel and posterior diagnostics rather than manually tuning every fit.

\subsection{Scope and future directions}\label{sec:limitations}

The scope of the present work should be kept clear. The mtanh parametrisation is, by construction, poorly suited to profiles with strong internal transport barriers or with multiple pedestal like features; for such cases a GP or spline parametrisation would be more appropriate. GP regression has already been shown to provide useful uncertainty quantification for fusion profile fitting and transport oriented uncertainty propagation \cite{chilenski2015gpr}, and more recent work has demonstrated regime flexible GP profile models with robust likelihoods for profiles containing outliers \cite{leddy2022singlegp}. A natural extension of the present workflow is therefore to replace or supplement the fixed mtanh parametrisation with a GP based profile module while retaining the same operational ideas: MDSplus based batch processing, robust likelihoods, posterior uncertainty estimates, and data quality diagnostics for each channel. This type of automated quality monitoring may become increasingly useful as KSTAR operating conditions evolve, for example under future tungsten wall operation, where diagnostic conditions and data quality patterns may change.

The present implementation uses the emcee affine-invariant ensemble sampler, which handles the elongated degeneracies of the mtanh cost surface well but may be less efficient than gradient-based methods such as Hamiltonian Monte Carlo for higher dimensional extensions of the model. The equilibrium reconstruction is taken as an input rather than inferred jointly with the kinetic profiles; coupling the present approach to a kinetic-EFIT loop with internal consistency checks is a natural next step.

The examples shown here demonstrate operational feasibility, diagnostic quality monitoring, and computational scaling relevant to large-scale use rather than a full database-level physics validation of all fitted pedestal parameters. Aggregate validation against independent diagnostics and downstream reconstruction impact are left for future work.

\section{Conclusions}\label{sec:conclusions}

We have presented an automated Bayesian profile fitting approach for KSTAR that mitigates two principal obstacles to automated kinetic profile reconstruction: contaminated diagnostic measurements and secondary minima on the multimodal mtanh cost surface.

The outlier problem is handled at the level of the likelihood. In the deployed workflow, the good-and-bad Gaussian mixture is the default likelihood for fitted diagnostic channels; fixed configurable mixture parameters yield posterior outlier probabilities for each channel as a quantitative data quality diagnostic. Gaussian and Sivia conservative likelihoods remain available as alternatives selected by users for sensitivity checks.

The local minimum problem is addressed by replacing the single point optimiser output with posterior exploration via MCMC. The emcee affine-invariant ensemble sampler \cite{foremanmackey2013emcee}, initialised from a MAP warm start, can reveal multimodal structure on the mtanh cost surface when the relevant basins are sampled and can produce posterior means and credible intervals that integrate over the accessible regions of significant probability, rather than conditioning on a single possibly secondary optimum.

The combination of robust likelihoods and MCMC exploration enables an end-to-end automation pipeline: the batch frontend processes an entire shot in a single command, and the MDSplus upload pipeline makes fitted profiles available for downstream KSTAR workflows. This capability is intended to support future large-scale KSTAR kinetic profile production for kinetic-EFIT reconstruction, TRANSP integrated modelling, and machine learning data set construction. The statistical structure is transferable to other tokamaks, although diagnostic observation models and data access layers would need to be adapted, and the combination of mixture-based robust likelihoods with MCMC sampling is a generally useful pattern for Bayesian inference problems with multiple diagnostics in which both data quality and cost surface convexity cannot be guaranteed.

\appendix
\setcounter{figure}{0}
\renewcommand{\thefigure}{A.\arabic{figure}}
\renewcommand{\theHfigure}{appendix.A.\arabic{figure}}

\section{Graphical interface and interactive quality control}\label{app:gui}

For inspection of individual time slices, the workflow provides a Qt graphical interface that shares the same fitting backend as the batch command line workflow. The graphical interface is not a separate analysis method; it is an interactive front end for loading the same diagnostic data, applying the same mtanh model, running the same Bayesian sampler, and inspecting the resulting profiles before or after automated batch processing. The interface is organised around four profile tabs corresponding to \(n_e\), \(T_e\), \(T_i\), and \(v_T\), with a common layout for shot and equilibrium selection, time navigation, fit execution, profile plotting, and display controls.

The plot area displays the enabled diagnostic points, uncertainty bars, disabled or rejected points, and the fitted profile summaries produced by the Bayesian workflow. Depending on the selected display mode, the user can compare the MAP profile, the posterior mean profile, and the posterior uncertainty band. When the Gaussian good-and-bad mixture likelihood is used, measurements with large posterior outlier probability can also be highlighted on the plot. The graphical display is therefore complementary to the outlier probability maps in the main text: the maps summarise repeated behaviour across many time slices, whereas the GUI allows individual slices to be checked in detail.

The interface is useful both before and after batch processing. Before a large run, representative time slices can be loaded to verify the diagnostic source, equilibrium mapping, channel masks, likelihood settings, and approximate profile behaviour. After a batch run, fitted results can be reloaded to inspect selected slices with unusual posterior summaries or high outlier probabilities. This interactive review step reduces the risk of launching or accepting an unattended run with an incorrect configuration, while keeping the statistical model and numerical backend identical to those used in the automated workflow.

\begin{figure*}[t]
\centering
\includegraphics[width=0.96\textwidth,height=0.78\textheight,keepaspectratio]{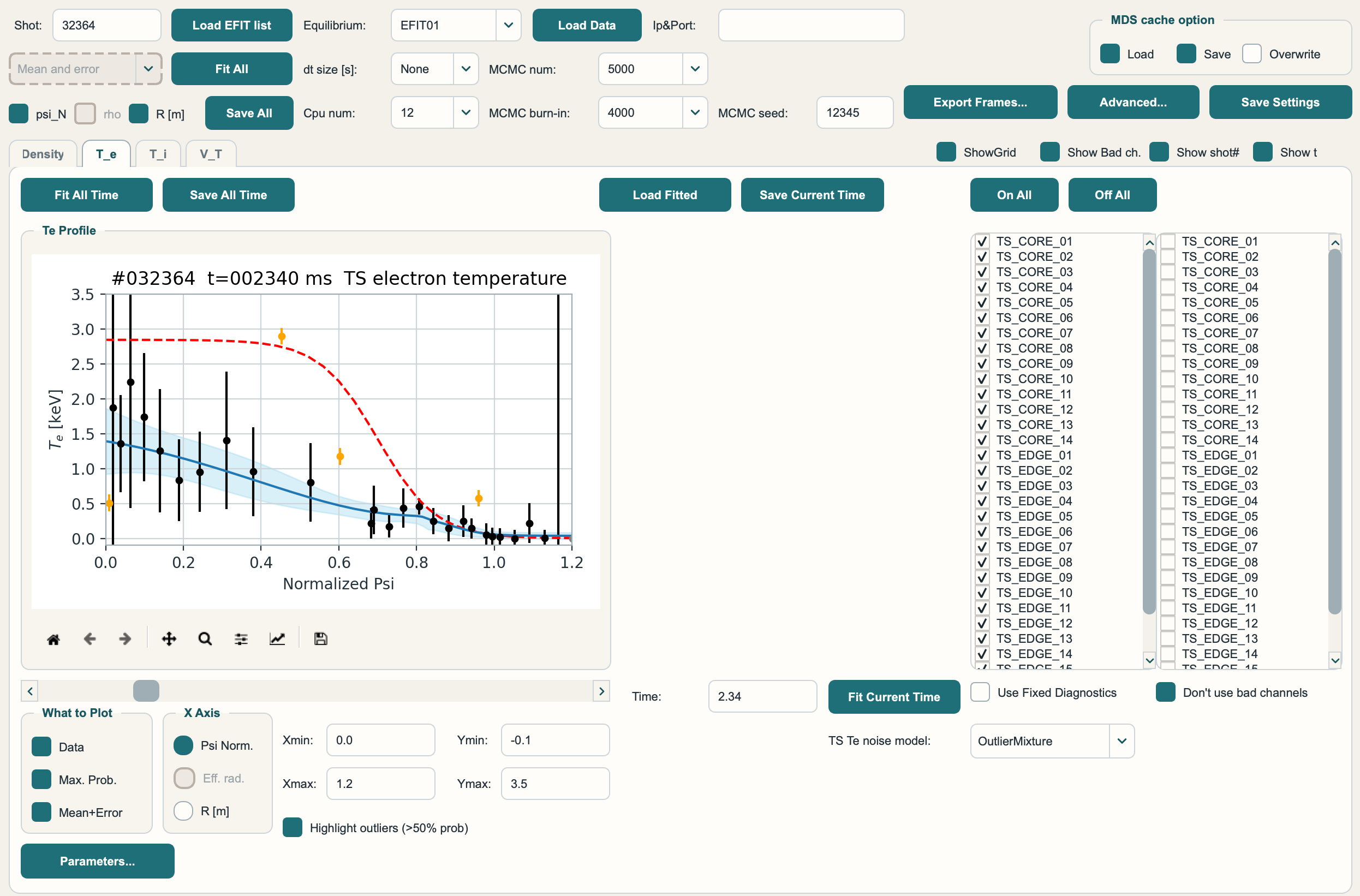}
\caption{Graphical interface for the \(T_e\) fit of KSTAR shot 32364 at \(t=2.34\,\mathrm{s}\) with the mixture likelihood. The GUI provides an interactive front end to the same fitting backend used by the batch workflow, allowing users to inspect diagnostic points, fitted profile summaries, uncertainty bands, and highlighted outlier candidates for individual time slices.}
\label{fig:gui}
\end{figure*}

\section*{Acknowledgements}

This research was supported by the R\&D Program of ``High Performance Tokamak Plasma Research \& Development (EN2601-17)'' and ``Korea-US Collaboration Research for High Performance Plasma on Tungsten Divertor (EN2603-02)'' through the Korea Institute of Fusion Energy (KFE) funded by the Government funds, Republic of Korea.

\bibliographystyle{unsrt}
\bibliography{biprofile_refs}

\begin{thebibliography}{10}

\bibitem{lao2005efit}
L.~L. Lao, H.~E. St.~John, Q.~Peng, J.~R. Ferron, E.~J. Strait, T.~S. Taylor,
  W.~H. Meyer, C.~Zhang, and K.~I. You.
\newblock {MHD} equilibrium reconstruction in the {DIII-D} tokamak.
\newblock {\em Fusion Science and Technology}, 48:968--977, 2005.
\newblock \doi{10.13182/FST48-968}.

\bibitem{pankin2025transp}
A.~Y. Pankin, J.~Breslau, M.~Gorelenkova, R.~Andre, B.~Grierson, J.~Sachdev,
  M.~Goliyad, and G.~Perumpilly.
\newblock {TRANSP} integrated modeling code for interpretive and predictive
  analysis of tokamak plasmas.
\newblock {\em Computer Physics Communications}, 312:109611, 2025.
\newblock \doi{10.1016/j.cpc.2025.109611}.

\bibitem{park2018fastran}
J.~M. Park, J.~R. Ferron, C.~T. Holcomb, R.~J. Buttery, W.~M. Solomon, D.~B.
  Batchelor, W.~Elwasif, D.~L. Green, K.~Kim, O.~Meneghini, M.~Murakami, and
  P.~B. Snyder.
\newblock Integrated modeling of high {$\beta_N$} steady-state scenario on
  {DIII-D}.
\newblock {\em Physics of Plasmas}, 25:012506, 2018.
\newblock \doi{10.1063/1.5013021}.

\bibitem{snyder2011eped}
P.~B. Snyder, R.~J. Groebner, J.~W. Hughes, T.~H. Osborne, M.~Beurskens, A.~W.
  Leonard, H.~R. Wilson, and X.~Q. Xu.
\newblock A first-principles predictive model of the pedestal height and width:
  development, testing and {ITER} optimization with the {EPED} model.
\newblock {\em Nuclear Fusion}, 51:103016, 2011.
\newblock \doi{10.1088/0029-5515/51/10/103016}.

\bibitem{lee2001kstar}
G.~S. Lee, M.~Kwon, C.~J. Doh, B.~G. Hong, K.~Kim, M.~H. Cho, W.~Namkung, C.~S.
  Chang, Y.~C. Kim, J.~Y. Kim, H.~G. Jhang, D.~K. Lee, K.~I. You, J.~H. Han,
  M.~C. Kyum, J.~W. Choi, J.~Hong, W.~C. Kim, B.~C. Kim, J.~H. Choi, S.~H. Seo,
  H.~K. Na, H.~G. Lee, S.~G. Lee, S.~J. Yoo, B.~J. Lee, Y.~S. Jung, J.~G. Bak,
  H.~L. Yang, S.~Y. Cho, K.~H. Im, N.~I. Hur, I.~K. Yoo, J.~W. Sa, K.~H. Hong,
  G.~H. Kim, B.~J. Yoo, H.~C. Ri, Y.~K. Oh, Y.~S. Kim, C.~H. Choi, D.~L. Kim,
  Y.~M. Park, K.~W. Cho, T.~H. Ha, S.~M. Hwang, Y.~J. Kim, S.~Baang, S.~I. Lee,
  H.~Y. Chang, W.~Choe, S.~G. Jeong, S.~S. Oh, H.~J. Lee, B.~H. Oh, B.~H. Choi,
  C.~K. Hwang, S.~R. In, S.~H. Jeong, I.~S. Ko, Y.~S. Bae, H.~S. Kang, J.~B.
  Kim, H.~J. Ahn, D.~S. Kim, C.~H. Choi, J.~H. Lee, Y.~W. Lee, Y.~S. Hwang,
  S.~H. Hong, K.~H. Chung, and D.~I. Choi.
\newblock Design and construction of the {KSTAR} tokamak.
\newblock {\em Nuclear Fusion}, 41:1515--1523, 2001.
\newblock \doi{10.1088/0029-5515/41/10/318}.

\bibitem{lee2010ts}
J.~H. Lee, S.~T. Oh, and H.~M. Wi.
\newblock Development of {KSTAR} {Thomson} scattering system.
\newblock {\em Review of Scientific Instruments}, 81:10D528, 2010.
\newblock \doi{10.1063/1.3494275}.

\bibitem{lee2016tci}
K.~C. Lee, J.-W. Juhn, Y.~U. Nam, Y.~S. Kim, H.~M. Wi, S.~W. Kim, and Y.-C.
  Ghim.
\newblock The design of two color interferometer system for the 3-dimensional
  analysis of plasma density evolution on {KSTAR}.
\newblock {\em Fusion Engineering and Design}, 113:87--91, 2016.
\newblock \doi{10.1016/j.fusengdes.2016.10.008}.

\bibitem{ko2010ces}
W.~H. Ko, H.~Lee, D.~Seo, and M.~Kwon.
\newblock Charge exchange spectroscopy system calibration for ion temperature
  measurement in {KSTAR}.
\newblock {\em Review of Scientific Instruments}, 81:10D740, 2010.
\newblock \doi{10.1063/1.3496991}.

\bibitem{sivia1996duff}
D.~S. Sivia.
\newblock Dealing with duff data.
\newblock In M.~Sears, V.~Nedeljkovic, N.~E. Pendock, and S.~Sibisi, editors,
  {\em Proceedings of the Maximum Entropy Conference}, pages 131--137, Port
  Elizabeth, South Africa, 1996. NMB Printers.

\bibitem{box1968outlier}
G.~E.~P. Box and G.~C. Tiao.
\newblock A {Bayesian} approach to some outlier problems.
\newblock {\em Biometrika}, 55:119--129, 1968.
\newblock \doi{10.1093/biomet/55.1.119}.

\bibitem{kim2024svmrgpr}
Minseok Kim, W.~H. Ko, Sehyun Kwak, Semin Joung, Wonjun Lee, B.~Kim, D.~Kim,
  J.~H. Lee, Choongki Sung, Yong~Su Na, and Y.~C. Ghim.
\newblock Kinetic profile inference with outlier detection using support vector
  machine regression and {Gaussian} process regression.
\newblock {\em Nuclear Fusion}, 64(10):106052, 2024.
\newblock \doi{10.1088/1741-4326/ad7304}.

\bibitem{krychowiak2004zeff}
M.~Krychowiak, R.~K{\"o}nig, T.~Klinger, and R.~Fischer.
\newblock {Bayesian} analysis of the effective charge from spectroscopic
  bremsstrahlung measurement in fusion plasmas.
\newblock {\em Journal of Applied Physics}, 96:4784--4792, 2004.
\newblock \doi{10.1063/1.1787135}.

\bibitem{kwak2026zeff}
S.~Kwak, M.~Krychowiak, M.~De~Bock, S.~Serov, and J.~Svensson.
\newblock {Bayesian} modelling for the visible spectroscopy reference system at
  {ITER}.
\newblock {\em Nuclear Fusion}, 66:056021, 2026.
\newblock \doi{10.1088/1741-4326/ae5722}.

\bibitem{stillerman1997mdsplus}
J.~A. Stillerman, T.~W. Fredian, K.~A. Klare, and G.~Manduchi.
\newblock {MDSplus} data acquisition system.
\newblock {\em Review of Scientific Instruments}, 68:939--942, 1997.
\newblock \doi{10.1063/1.1147719}.

\bibitem{groebner2001mtanh}
R.~J. Groebner, D.~R. Baker, K.~H. Burrell, T.~N. Carlstrom, J.~R. Ferron,
  P.~Gohil, L.~L. Lao, T.~H. Osborne, D.~M. Thomas, W.~P. West, J.~A. Boedo,
  R.~A. Moyer, G.~R. McKee, R.~D. Deranian, E.~J. Doyle, C.~L. Rettig, T.~L.
  Rhodes, and J.~C. Rost.
\newblock Progress in quantifying the edge physics of the {H} mode regime in
  {DIII-D}.
\newblock {\em Nuclear Fusion}, 41:1789--1802, 2001.
\newblock \doi{10.1088/0029-5515/41/12/306}.

\bibitem{lee2026nnces}
J.~K. Lee, W.~H. Ko, H.~H. Lee, B.~Kim, Y.~H. Lee, G.~W. Shin, J.~Kim, and
  J.~M. Lee.
\newblock Neural network-based analysis of charge exchange spectra in {KSTAR}.
\newblock {\em Fusion Engineering and Design}, 222:115518, 2026.
\newblock \doi{10.1016/j.fusengdes.2025.115518}.

\bibitem{nelder1965simplex}
J.~A. Nelder and R.~Mead.
\newblock A simplex method for function minimization.
\newblock {\em The Computer Journal}, 7:308--313, 1965.
\newblock \doi{10.1093/comjnl/7.4.308}.

\bibitem{goodman2010ensemble}
J.~Goodman and J.~Weare.
\newblock Ensemble samplers with affine invariance.
\newblock {\em Communications in Applied Mathematics and Computational
  Science}, 5:65--80, 2010.
\newblock \doi{10.2140/camcos.2010.5.65}.

\bibitem{foremanmackey2013emcee}
D.~Foreman-Mackey, D.~W. Hogg, D.~Lang, and J.~Goodman.
\newblock emcee: the {MCMC} hammer.
\newblock {\em Publications of the Astronomical Society of the Pacific},
  125:306--312, 2013.
\newblock \doi{10.1086/670067}.

\bibitem{lee2026prism}
J.~K. Lee.
\newblock An integrated multi-diagnostic visualization platform for {KSTAR}
  tokamak.
\newblock {\em Fusion Engineering and Design}, 228:115786, 2026.
\newblock \doi{10.1016/j.fusengdes.2026.115786}.

\bibitem{chilenski2015gpr}
M.~A. Chilenski, M.~Greenwald, Y.~Marzouk, N.~T. Howard, A.~E. White, J.~E.
  Rice, and J.~R. Walk.
\newblock Improved profile fitting and quantification of uncertainty in
  experimental measurements of impurity transport coefficients using {Gaussian}
  process regression.
\newblock {\em Nuclear Fusion}, 55:023012, 2015.
\newblock \doi{10.1088/0029-5515/55/2/023012}.

\bibitem{leddy2022singlegp}
J.~Leddy, S.~Madireddy, E.~Howell, and S.~Kruger.
\newblock Single {Gaussian} process method for arbitrary tokamak regimes with a
  statistical analysis.
\newblock {\em Plasma Physics and Controlled Fusion}, 64:104005, 2022.
\newblock \doi{10.1088/1361-6587/ac8b50}.

\end{thebibliography}

\end{document}